# Quantum coherent control of self-induced ultraslow light

B. S. Ham[*] and J. S. Han

*Center for Photon Information Processing, and the Graduate School of Information and Telecommunications, Inha University, 253 Yonghyun-dong, Nam-gu, Incheon 402-751, S. Korea*
*Corresponding author: bham@inha.ac.kr*

Quantum coherent control of slow light for all-optical switching is investigated in a multi-level system of solids for an understanding of self-induced ultraslow light. In an optical population shelving system of a rare-earth doped solid, dynamics of the slow light are presented by using a third optical field controlling shelved atom population. Unlike two-photon coherence-based delayed all-optical switching utilizing electromagnetically induced transparency, the present method relies on one-photon coherence controlling shelved atom population.

Ultraslow group velocity [1,2] in inhomogeneously broadened solids has been studied for potential application to quantum optical information processing such as quantum memory [1,3], enhanced nonlinear optical process [4], all-optical buffer memory [5], and quantum entanglement generation [6]. For the control of group velocity of traveling light several techniques have been demonstrated using methods such as electromagnetically induced transparency (EIT) [1,3,4,6], coherent population oscillation [2], and a spectral hole-burning technique [2,5]. Both EIT and coherent population oscillation techniques need extra light to couple the system to drive refractive index modification. The spectral hole-burning technique is relatively simple because it uses the same frequency light, but is limited in controlling group delay time. Here we present dynamics of hole-burning based slow light for all-optical switching applications, as well as fundamental physics of the dynamic control in an atom shelved model.

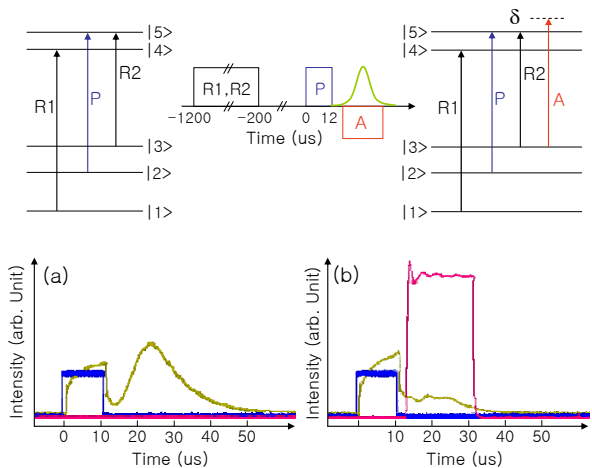

**Fig. 1.** Quantum coherent control of self-induced slow light. (a) Self-induced ultraslow light, and (b) All-optical switching of the self-induced slow light. Inset: Schematic energy level diagram of Pr:SYO and pulse sequence. The intensity of R1, R2, A, and P are 25, 11, 10, and 3 W/cm$^2$, respectively.

Figure 1 shows quantum coherent control of ultraslow light for all-optical switching in a rare earth $Pr^{3+}$ doped $Y_2SiO_5$ (Pr:YSO) at T = 5 K. Figure 1(a) shows self-induced ultraslow light in a two-level system (see the energy level diagram interacting with four laser pulses (R1, R2, P, and A) based on persistent spectral hole-burning [7]. All laser beams are not collinear, but overlap inside the crystal in about 80% one another. In the energy-level diagram of Fig. 1, the R1 and R2 act as repump pulses used for (re)initialization of the system. Ideally the normalized population in state |2> is initially ~1/3 due to thermally stabilized Boltzman distribution in the hyperfine states of Pr:YSO. By the action of the repump, the spectrally chosen population in state |2> can be increased up to three times. According to the group velocity theory, the more atoms are related, the slower group velocity is reached [8]. The weak probe pulse P burns a hole resulting in narrow absorption spectrum modification, where the spectral width of the transparency window of the probe in the absorption spectrum is determined by the laser jitter [9]. The laser jitter of the probe pulse P is as narrow as 300 kHz, which is comparable with the transparency window obtained by EIT [10]. The result of the narrow transparency window in state |2> brings a very stiff dispersion slope resulting in self-induced ultraslow light. In Fig. 1(a), the yellow-green (blue) line represents the probe P (electronic pulse) followed by the slow light.

In the pulse sequence shown in the inset of Fig. 1, the front part of the probe P must be absorbed by the dense medium resulting in atom excitation into state |5> burning a narrow spectral hole, then the rest of P experiences group delay. When the control pulse A follows P, the control A resonantly depopulates state |5> into state |3>. This population control of state |5> induces coherence modification of the probe resulting in absorption change (one-photon coherence, Im$\rho_{13}$).

In Fig. 1(b) all-optical switching of the self-induced ultraslow light is observed by applying the control pulse A (red line, $\delta = 0$). Here it is clear that there is no coherence among the pulses R1, R2, and P, because the optical coherence cannot be sustained longer than the inverse of the laser jitter, which is ~5 μs (the pulse separation is 200 μs). Moreover, the 200 μs delay between the repump and the probe is long enough for the excited atoms to decay down to the ground states (optical $T_1$ = 164 μs in Pr:YSO) [9]. Unlike the delayed all-optical switching using ultraslow light based on EIT [11], where atomic coherence (two-photon coherence Re$\rho_{12}$) plays a major role, Fig. 1(b) demonstrates that the self-induced ultraslow light can also be switched simply by controlling the excited state population (discussed in more detail in Fig. 3). The all-optical switching in Fig. 1(b) has potential for postselective all-optical processing, when pre-activated particular optical data is required to be purged in an optical pulse stream.

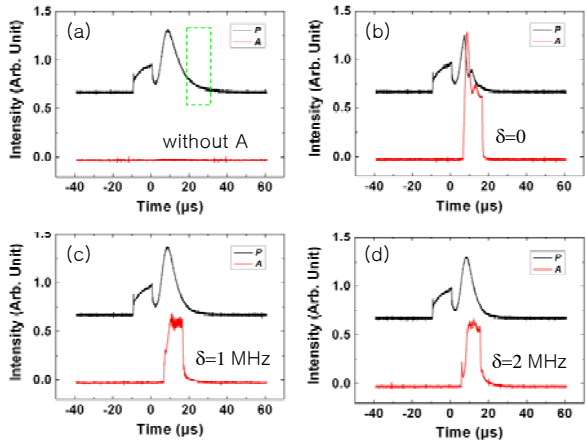

**Fig. 2.** (a) Self-induced ultraslow light, (b)-(d) all-optical switching with control A for (b) δ=0, for (c) δ=1 MHz, and for (d) δ= 2 MHz. The intensity of A is 12 W/cm²

Figure 2 represents all-optical switching of the self-induced ultraslow light with detuning δ of the control pulse A. The control A (dotted green box) is designed to be turned on at the peak intensity position of the slow light (black line) as shown in Fig. 2(a). When an optimum condition is met, both A (δ=0, red line) and the slow light oscillate as shown in Fig. 2(b). This transient behavior will be discussed in more detail in Fig. 3. If the control A is detuned (δ≠0) to not depopulate the excited atoms, then the slow light is free from switching as shown in Figs. 2(c) and 2(d). No optical coherence exists between the slow light and the control A, because of the long delay between them.

For Fig. 3, the power of the control A used in Fig. 1(b) is increased. As shown in Fig. 3(a) both the control and the slow light simultaneously experience transient behavior: Damped Rabi flopping. To analyze this behavior, we have numerically calculated Fig. 3(a) in Fig. 3(b) by solving density matrix equations of motion [11]. As shown in Fig. 3(b), both population difference ($\rho_{55}-\rho_{33}$; $|5\rangle - |3\rangle$) and the slow light intensity oscillate in a way very similar to the resonant control pulse A. Here it should be noted that the initial population in state $|3\rangle$ is almost zero for the A, as mentioned above in Fig. 1. Thus, the control A triggers the excited atoms in state $|5\rangle$ and then transfers them into state $|3\rangle$. If the pulse area of the control A is more than π, the transferred atoms in state $|3\rangle$ must be returned back to state $|5\rangle$. Because this kind of Rabi flopping (population oscillation in state $|5\rangle$) by the pulse A is damped, one can conclude that the oscillation amplitude must be decreased as time goes on. As shown in Fig. 3(b), the transient dynamics of the slow light (intensity modulation) are caused by the population change in the excited state $|5\rangle$.

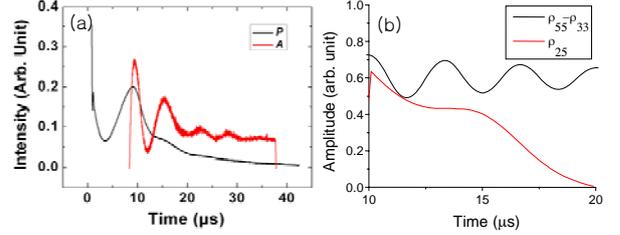

**Fig. 3.** (a) Transient behavior of the all-optical switching based on Rabi flopping of the population. (b) Numerical calculations: $G_{25}=G_{35}=1$ kHz, $g_{25}=g_{35}=50$ kHz, $\Omega_P=10$ kHz, $\Omega_A=100$ kHz. The intensity of P and A are 3 and 10 W/cm².

In the numerical calculations, we assumed a Gaussian pulsed slow light whose pulse length (FWHM) is 10 μs centered at the pulse A. For the one-photon coherence Im$\rho_{13}$, photon loss (intensity decrease) of the slow light is caused by absorption (negative values of the coherence Im$\rho_{13}$) due to a population decrease in state $|5\rangle$. If the population in the excited state $|5\rangle$ is modified by the action of the control A, then the initially established balance between states $|2\rangle$ and $|5\rangle$ should be broken. Thus, the slow light begins to be absorbed until the negative value of the coherence (Im$\rho_{13}$) becomes positive. This turning point is met whenever the slope of the population oscillation for $|5\rangle - |3\rangle$ reaches a minimum value. Therefore, the population decrease in the excited state $|5\rangle$ by the control A induces photon absorption from the slow light, so that the slow light intensity becomes decreased as shown in Fig. 3(a) (see the black line). As shown in Fig. 3, the control pulse-induced population change leads to the dynamics of the slow light.

Figure 4 presents slow light modulations with different intensities of the control A. The oscillation frequency of the slow light in Fig. 4(a) increases as the intensity of the pulse A increases. Fig. 4(b) shows the oscillation frequency is linearly proportional to the square root of intensity (Rabi frequency) of the control A. This proves that the atom population change in the excited state $|5\rangle$ causes the slow light intensity modulation. Thus, a complete elimination of the slow light must be possible if one can remove the atoms in the excited state $|5\rangle$ permanently. One easy way to do this is to apply a π pulse of the control A, where the population inversion between states |5> and |3> is induced.

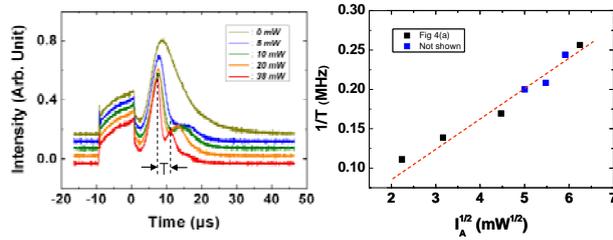

**Fig. 4.** (a) Transient effect of the probe P. The maximum intensity of A is same as that in Fig. 2(b). (b) Oscillation frequency of P versus square root of A power.

In conclusion we demonstrated quantum coherent control of self-induced ultraslow light in a populating shelving system of a rare-earth doped crystal for understanding fundamental physics as well as for potential applications to slow-light-based all-optical switching. Unlike EIT-based delayed all-optical switching using two-photon coherence [12], the present method has the unique property of controlling excited population by using a third optical pulse (one-photon coherence). Here, the third optical pulse need not be coherent with the slow light. The all-optical switching using slow light is important for post selective control in an optical pulse stream for ultrahigh-speed all-optical information processing.

**Acknowledgement**
This work was supported by the CRI program (Center for Photon Information Processing) of MEST via KOSEF, S. Korea.